\documentstyle[12pt,epsf]{article}
\begin{document}
\renewcommand{\thefootnote}{\fnsymbol{footnote}}
\begin{flushright}
KEK-preprint-94-109\\
NWU-HEP 94-05\\
TUAT-HEP 94-06\\
DPNU-94-41\\
TIT-HPE 94-08\\
OCU-HEP 94-08\\
PU-94-687\\
INS-REP-1063\\
KOBE-HEP 94-07\\
\end{flushright}
\vskip -3cm
\epsfysize3cm
\epsfbox{kekm.epsf}
\begin{flushleft}
{\bf \Large Measurement of inclusive electron
 cross section in $\gamma \gamma$ collisions at TRISTAN
\footnote{To be published in Phys. Lett. {\bf B}.}
}
\vskip 0.5cm
\underline{
M.Iwasaki$^{(1)}$}
\footnote{internet address: masako@kekvax.kek.jp.},
R.Enomoto$^{(2)}$,
H.Hayashii$^{(1)}$,
E.Nakano$^{(3)}$,
K.Abe$^{(3)}$,
T.Abe$^{(3)}$,
I.Adachi$^{(2)}$,
K.Adachi$^{(1)}$,
M.Aoki$^{(3)}$,
M.Aoki$^{(4)}$,
S.Awa$^{(1)}$
K.Emi$^{(5)}$,
H.Fujii$^{(2)}$,
K.Fujii$^{(2)}$,
T.Fujii$^{(6)}$,
J.Fujimoto$^{(2)}$,
K.Fujita$^{(7)}$,
N.Fujiwara$^{(1)}$,
B.Howell$^{(8)}$,
N.Iida$^{(2)}$,
H.Ikeda$^{(2)}$,
Y.Inoue$^{(7)}$,
R.Itoh$^{(2)}$,
H.Iwasaki$^{(2)}$,
R.Kajikawa$^{(3)}$,
K.Kaneyuki$^{(4)}$,
S.Kato$^{(9)}$,
S.Kawabata$^{(2)}$,
H.Kichimi$^{(2)}$,
M.Kobayashi$^{(2)}$,
D.Koltick$^{(8)}$,
I.Levine$^{(8)}$,
S.Minami$^{(4)}$,
K.Miyabayashi$^{(3)}$,
A.Miyamoto$^{(2)}$,
K.Muramatsu$^{(1)}$,
K.Nagai$^{(10)}$,
K.Nakabayashi$^{(3)}$,
O.Nitoh$^{(5)}$,
S.Noguchi$^{(1)}$,
A.Ochi$^{(4)}$,
F.Ochiai$^{(11)}$,
N.Ohishi$^{(3)}$,
Y.Ohnishi$^{(3)}$,
Y.Ohshima$^{(4)}$,
H.Okuno$^{(9)}$,
T.Okusawa$^{(7)}$,
T.Shinohara$^{(5)}$,
A.Sugiyama$^{(3)}$,
S.Suzuki$^{(3)}$,
S.Suzuki$^{(4)}$,
K.Takahashi$^{(5)}$,
T.Takahashi$^{(7)}$,
T.Tanimori$^{(4)}$,
T.Tauchi$^{(2)}$,
Y.Teramoto$^{(7)}$,
N.Toomi$^{(1)}$,
T.Tsukamoto$^{(2)}$,
T.Tsumura$^{(5)}$,
S.Uno$^{(2)}$,
T.Watanabe$^{(4)}$,
Y.Watanabe$^{(4)}$,
A.Yamaguchi$^{(1)}$,
A.Yamamoto$^{(2)}$, and
M.Yamauchi$^{(2)}$\\
\vskip 0.5cm
(TOPAZ Collaboration)\\
\vskip 0.5cm
{\it
$^{(1)}$Department of Physics, Nara Women's University,
 Nara 630, Japan
\\
$^{(2)}$National Laboratory for High Energy Physics, KEK,
  Ibaraki-ken 305, Japan
\\
$^{(3)}$Department of Physics, Nagoya University,
 Nagoya 464, Japan
\\
$^{(4)}$Department of Physics, Tokyo Institute of Technology,
     Tokyo 152, Japan
\\
$^{(5)}$Dept. of Applied Physics,
Tokyo Univ. of Agriculture and Technology,
 Tokyo 184, Japan
\\
$^{(6)}$Department of Physics, University of Tokyo,
   Tokyo 113, Japan
\\
$^{(7)}$Department of Physics, Osaka City University,
 Osaka 558, Japan
\\
$^{(8)}$Department of Physics, Purdue University,
 West Lafayette, IN 47907, USA
\\
$^{(9)}$Institute for Nuclear Study, University of Tokyo,
   Tokyo 188, Japan
\\
$^{(10)}$The Graduate School of Science and Technology,
Kobe University,
Kobe 657, Japan
\\
$^{(11)}$Faculty of Liberal Arts, Tezukayama Gakuin University,
 Nara 631, Japan
}
\end{flushleft}
\newpage
\begin{abstract}
\baselineskip = 20pt
We have studied open charm production in $\gamma \gamma$ collisions
with the TOPAZ detector at the TRISTAN $e^{+}e^{-}$ collider.
In this study, charm quarks were identified by
electrons (and positrons) from semi-leptonic decays of charmed
hadrons.
The data corresponded to an integrated luminosity of 95.3 pb$^{-1}$
at a center-of-mass energy of 58 GeV.
The results are presented as the cross sections of inclusive electron
production in $\gamma \gamma$ collisions with an anti-tag condition,
as well as the subprocess cross sections, which correspond
to resolved-photon processes.
The latter were measured by using
a sub-sample with remnant jets.
A comparison with various  theoretical
predictions based on  direct and resolved-photon processes
showed that our data prefer that with relatively large gluon
contents in a photon at small $x (x \le 0.1)$, with the
next-to-leading order correction, and with a charm-quark mass of 1.3 GeV.

\end {abstract}
\newpage
\baselineskip = 24pt
\section{Introduction}
Recently,
several experimental collaborations at PEP/PETRA\cite{PEPPETRA},
 TRISTAN\cite{HAYASHII,AMY}, and LEP\cite{ALEPH,DELPHI}
have reported that high-transverse momentum($P_{T}$) jet
production  in   (quasi-)real $\gamma \gamma$ collisions
 requires a new mechanism (resolved-photon process)
 in addition to the direct photon process
 $\gamma \gamma \rightarrow q \bar{q}$(Fig.\ref{fig1}-(a)).
The resolved-photon processes are such mechanisms in which the
quark-gluon contents of photons participate in hard scattering
to cause the high-$P_{T}$ jet production[6-8].
 The relevant diagrams are
shown in Figs. \ref{fig1}-(b) and (c) for one- and two-resolved-photon
processes, respectively.

The analyses of $\gamma \gamma$ collisions at TRISTAN\cite{HAYASHII}
and  $\gamma p$ collisions at HERA\cite{H1ZEUS}
have provided clear results for the production of remnant jets
which are regarded as the evidence of the resolved-photon processes.

These recent data on jet production have provided us with
some important information concerning the quark and gluon contents of a
photon: they are inconsistent with a model which has very hard gluon
distribution (LAC3\cite{LAC}).
In the jet analysis\cite{HAYASHII}, however, the sensitivity of the data was
 limited  to high $x$ values $( x > 0.1)$ because of the
experimental $P_T$ cut
applied in jet reconstruction and a theoretical cut ($P_T^{min}$)
introduced in order to justify the perturbative calculation.

 Charm-quark production in $\gamma \gamma$ collisions is
a good tool for studying the jet production mechanisms
in two-photon processes\cite{NLO}.
   The reaction has several advantages over light-quark production.
Since the charm-quark mass is (relatively) well defined,
the  free parameter $P_T^{min}$ is unnecessary.
The dominant mechanisms for
$c\bar{c}$ production are only the direct and
 the one-resolved processes(photon-gluon fusion);
the contribution from the two-resolved process is small.
Charmed-hadron production from
the vector-meson-dominance model(VDM: Fig.\ref{fig1}-(d))\cite{VDM} is also
expected to be negligible in the TRISTAN energy region.
It should also be emphasized that higher order QCD corrections
(O($\alpha_{s}$))
are available for both the direct and  one-resolved
processes\cite{NLO,enomotoprd,enomotopl}.
Studies of charm production in $\gamma
\gamma$ collisions thus provide a good opportunity to probe the gluon
density in the photon.

In this paper we report on a measurement of open charm
production in $\gamma \gamma$ collisions under an anti(no)-tag condition.
In the PEP/PETRA experiments\cite{CPEP} and in our previous
paper\cite{enomotoprd,enomotopl}, charm quarks were identified
by detecting $D^{*\pm}$ signals. In this paper,  we use
electrons and positrons in the final states
to tag charm-quark production in two-photon processes
(hereafter, electron also implies positron).
The inclusive electron method is sensitive to
the charm-quark production at relatively low $P_T$,
where the one-resolved process contributes
significantly. We can access the small Bjorken-$x$ region down to
$x \sim 0.02$, where model dependence is the largest
among various parametrizations of the gluon
density in the photon[10,16-20].
The same kind of inclusive electron analysis was carried out
by the VENUS collaboration\cite{VENUS}.

\subsection{Cross section and Monte-Carlo simulation}
In order to compare the data with theoretical
predictions of the direct and the resolved processes,
we carried out a Monte-Calro simulation for each process.
The cross section for charm-quark pair production in the two-photon
process $ e^+ e^- \rightarrow e^+ e^- c \bar{c} X $ was obtained by
applying
 the equivalent real-photon approximation(EPA)  to the nearly
on-shell virtual photons emitted by the beam electrons\cite{EPA}.
Under the anti-tag condition imposed on the scattered electrons,
the photon flux factor can be written as\cite{AU93,HAG93,HAG91,flux}
\begin{eqnarray}
    f_{\gamma/e}( x_{\gamma})& =& \frac{ \alpha_{em}}{2 \pi x_{\gamma}}
                 \left( 1 + ( 1 - x_{\gamma})^{2} \right)
                 \ln \frac{P^{2}_{max}}{P^{2}_{min}}
           \     -  \ \frac{\alpha_{em}}{\pi}
\frac{1-x_{\gamma}}{x_{\gamma}},
\end{eqnarray}
where
\begin{eqnarray}
       P^{2}_{min} & = & m_e^{2} \frac{ x_{\gamma}^{2}}{1 - x_{\gamma}}.
\label{eqn:flux}
\end{eqnarray}
Here, $ P^{2}_{min}$ is the kinematical minimum of the photon
virtuality. $P^{2}_{max}$ is given by
\begin{equation}
 P^{2}_{max} = min( {P^{2}_{max,kin}, Q^{2}}),
\end{equation}
where $P^{2}_{max,kin} = 2 E_{beam}^{2}(1 - x_{\gamma})( 1 -
\cos\theta_{max})$
is the maximum photon virtuality determined by the experimental
anti-tag condition.
For the scale $Q^{2}$, we set
$Q^{2} = m_{c}^{2} + P_{T}^{2}$\cite{HAG93}.
In this experiment, the scattering angles of the
beam electrons were limited to less than $3.2^{\circ}$
if their energies were
greater than 0.4 $E_{beam}$(i.e. $\theta_{max}$ = $3.2^{\circ}$ for
$x_{\gamma}=E_{\gamma}/E_{beam}<0.6$). $P^{2}_{max}$ was mainly
determined  by this experimental anti-tag condition.
To check the validity of the approximation,  the EPA prediction
was compared with the exact matrix element calculation\cite{KUR88}
for the direct $ e^+ e^- \rightarrow e^+ e^- c \bar{c}$ process. We
found that  both predictions agree at a 1\% level
under this anti-tag condition\cite{HAG93}.

For the parton distribution inside the photon, we have tried
two typical parametrizations given by Drees-Grassie(DG)\cite{DG} and
by Levy-Abramowicz-Charchula(LAC1)\cite{LAC} as working examples.
The gluon distribution of LAC1 increases more rapidly than that of
DG at $x$ less than 0.1.

For the cross sections of the subprocesses ($\gamma \gamma \rightarrow
c \bar{c}$, $\gamma g \rightarrow c \bar{c}$, $ g g \rightarrow c
\bar{c}$, and $q \bar{q} \rightarrow c \bar{c}$) we used the lowest
order formulas including the charm-quark mass. The higher order
effects were taken into account by appropriately weighting the
$P_{T}$ distribution of charm quarks\cite{NLO,enomotoprd,enomotopl}.
We used the BASES/SPRING program\cite{BASES} to
calculate the cross sections of these processes and to generate
events at the parton level.
 The charm-quark mass($m_{c}$) used in the cross-section calculations
was changed in the range
1.3 - 1.5 GeV in order to see the effect of the mass.
The threshold  was set at
 $2 \times  m_{D}= 3.74 $ GeV for the mass of the $\gamma \gamma$ system
 and at
2 $\times$ 1.6 GeV for that of the $c\bar{c}$ system.
The remnant jets in the resolved process were generated in the beam
direction.
The produced partons were hadronized by JETSET 6.3\cite{LUND}.
In this program,
the (constituent) charm-quark mass($m_{const.}$) was set to be 1.6
GeV. To ensure
 energy conservation, we adjusted the momentum of
the charm quark in the fragmentation stage
as $P_{JETSET}^{2} =
P_{gen}^{2} + m_{c}^{2} - m_{const.}^{2}$.
For the fragmentation of the charm quark,
we used the Peterson function, and set the parameter
$\epsilon_{c}$ to be 0.07.
The generated events were passed
through the TOPAZ detector simulator so as to take into
account any acceptance and resolution effects.

\section{Apparatus and event selection}
The data were taken with the TOPAZ detector at the TRISTAN
$e^{+}e^{-}$ collider.
The integrated luminosity
was 95.3 pb$^{-1}$ and the average center-of-mass energy was 58.0 GeV.
The details of the TOPAZ detector are described
in references\cite{TOPAZ, FCL}. In this analysis, we used a time-projection
chamber (TPC) for tracking and dE/dx measurements for charged
tracks. Plastic scintillation counters(TOF) surrounding the TPC
were used for time-of-flight measurements.
We detected electromagnetic showers with three kinds of calorimeters:
a barrel lead-glass calorimeter (BCL), an end-cap
Pb-proportional-wire-counter-sandwich calorimeter (ECL), and
a forward bismuth-germanate-crystal
 calorimeter (FCL), which cover polar angular ranges
$ | \cos{\theta} | \leq 0.85$,
$0.85 \leq | \cos{\theta} | \leq 0.98$ ,
and $0.972 \leq | \cos{\theta} | \leq 0.998$ $(=3.2^{\circ})$, respectively.
Since the FCL is very close to the beam pipe, the calorimeter is
protected from beam-induced background by an extensive shielding
system\cite{KI93}.

\subsection{Two-photon event selection}
Two-photon events were detected mainly by the charged-track trigger.
The trigger required at least
two charged tracks with $P_T$ $>$ 0.3 - 0.7 GeV
and an opening angle $>$ 45 - 70 degrees (depending on the beam
condition). On the other hand,
the neutral energy trigger required that the energy
deposit in the BCL had to be greater than 2 $\sim$ 4 GeV,
or that in the ECL had to be greater than 10 GeV.
The details concerning the trigger system can be
found in reference\cite{trig}.
The trigger efficiency for charmed events
 in two-photon processes was
 estimated with a trigger simulation to be about 93\%.

The hadronic events produced by two-photon interactions were selected
based on the following criteria:
\begin{enumerate}
\item The number of charged tracks with $P_T > 0.15$ GeV and the polar
      angle $|\cos\theta| < 0.83 $ had to be at least 4;
\item The position of the origin of the event(i.e. the event vertex),
      reconstructed from all tracks, had to be within 1.5 cm
      in the xy-plane and within $\pm$ 2.0 cm along the beam line
      from the interaction point;
\item The  visible energy($E_{vis}$) of the event
      had to satisfy  $E_{vis}\leq 30$ GeV,
      where both the charged tracks in the TPC and the neutral clusters
      in the BCL were
      used in the calculation of $E_{vis}$;
\item The  mass of the system of the observed hadrons($W_{vis}$) had to
      be  $W_{vis}\geq 3$ GeV, where the tracks in the TPC
      and the clusters in the BCL
      were used; and
\item The energy of the most energetic cluster appearing in the BCL,
      the ECL, or the FCL had to be less than $0.4 E_{beam}$.
\end{enumerate}

Criterion 5 ensures the anti-tag condition,
which limits the scattering angles of the beam electrons to less than
$3.2^{\circ}$. We call these events an anti-tag sample.
Substantial part of which had
some activities, though of relatively low energy,
in the FCL.  Our Monte-Calro study showed that these activities can be
naturally understood by remnant jets in the resolved-photon
processes\cite{HAYASHII,TAU94}.
In order to more closely study these events,
we selected them by adding the following criterion:
  the energy deposit in the FCL had to be
in the range 0.5 GeV $< E_{vis}^{FCL} < 0.25 E_{beam}$;
they are called remnant-jet-tagged events.
 These selection cuts left 27379 anti-tag and  13508
remnant-jet-tag events.

\subsection {Electron selection}
Among the selected events, electron-track candidates were searched in
the momentum range $0.4 \leq P_T \leq 3$ GeV and the polar angle rage
$ | \cos \theta | \leq 0.77$.

The TOPAZ detector allows three methods for electron identification.
 The energy loss
(dE/dx) information from the TPC enables us to separate
electrons from hadrons in the momentum region($P_T < 3$GeV)\cite{nagai}.
The E/P ratio of the energy(E) measured by the BCL
and the momentum(P) measured by the TPC can separate
electrons clearly.
The TOF was useful to resolve electrons from Kaons and protons in the
 overlapped region of dE/dx.

Since the largest background source in this analysis
was the electrons from the
$\gamma$-conversions at the material in front  of the TPC,
we first rejected the dominant part of such  electrons by the following
methods.
We reconstructed secondary vertices ($V^{0}$'s) from all
combinations of two tracks, and
calculated the invariant mass of each $V^{0}$ assuming that its
daughter particles are electrons.
For the $V^{0}$-reconstruction,  two
kinds of vertices, i.e.,  non-crossing and crossing cases in the xy-plane
(perpendicular to the beam axis) were searched.
In the former case, the distances of the two tracks
at the minimum distance position in the xy-plane were required
to be less than 7 cm in the xy-plane and 3 cm in the z-direction.
In the latter case, we chose from the two crossing points
that with the shorter z-difference, and required it to be
less than 1.5 cm.
We then rejected the tracks in the pair if its invariant mass was
 $\leq$ 80 MeV in the former or $\leq$ 150 MeV
in the latter cases.

We also required the closest approach of each electron-track candidate
to the event vertex in the xy-plane to be $<$ 0.5-1.5 cm depending on
$P_{T}$.

Among the remaining tracks, electron tracks were searched
by combining the information from the E/P ratio, dE/dx,
and TOF as follows:
(1) Charged tracks in the TPC were extrapolated to the BCL.
We then selected, for each TPC track, the BCL cluster
which was the closest. The E/P of each of the so-selected TPC-BCL
combinations had to satisfy $0.75 \leq$ E/P $\leq 1.25$.
(2) The dE/dx was calculated for the electron-track candidate
to be used in the subsequent electron counting.
The resultant dE/dx was mostly contained
in the range $5.5 \leq dE/dx \leq 7.5$ keV/cm.
(3) To remove Kaons and protons in the dE/dx overlap region,
we used information from the TOF and required the track to have
a confidence level(CL) of 0.01 or better for the electron hypothesis.

The performance of electron identification is demonstrated
in  Fig.\ref{fig:selection},
where various distributions
(the closest TPC-BCL distance(Fig.\ref{fig:selection}-(a)),
the E/P ratio(Fig.\ref{fig:selection}-(b)),
the CL for the electron hypothesis in the TOF(Fig.\ref{fig:selection}-(c)),
and the dE/dx(Fig.\ref{fig:selection}-(d))) are shown.
Notice that these figures were obtained
with the electron candidates selected by all of
the cuts but the one on each plotted quantity.
Notice in particular that Figs.\ref{fig:selection}-(a) to (c) were
made with an dE/dx cut ($5.5 \leq dE/dx \leq 7.5$ keV/cm) which was
used only for purpose of displaying.
 As can be seen from the dE/dX distribution in Fig.\ref{fig:selection}-(d),
two peaks corresponding to electrons and pions are clearly separated.

We counted the numbers of electrons in each $P_{T}$ bin by
fitting the dE/dx distributions with double Gaussians bin by bin.
The $P_T$ binning was selected so as to approximately equalize
the number of entries in each bin.
The $P_T$ resolutions were smaller than the bin width.
The numbers obtained are plotted (closed circles) in Fig.\ref{fig:bg}.
The errors are statistical.
The numbers of electron candidates summed over the $P_{T}$ bins were
 $214.8\pm 15.4$ and $88.0\pm 9.3$
for the anti-tag and the remnant-jet-tag samples, respectively.

\subsection{Background estimation}
These electron candidates included background tracks coming from
sources other than charm-quark decays. These background sources were
studied in detail.

There remained electrons coming  from $\gamma$-conversions
which escaped from pair reconstruction.
They were presumably energy-unbalanced pairs,
for each of which the lower $P_T$ track was not
reconstructed by  the TPC.
We estimated the failure rate of the $V^{0}$
reconstruction by a Monte-Carlo simulation.
The failure rate,
  $ \eta = N^{M.C.}_{V^{0}failure}/N^{M.C.}_{V^{0}reconstructed}$,
was estimated for each $P_T$-bin and was typically $\sim$0.4.
In the calculation, we included contributions from
 Dalitz decay($\pi^{0} \rightarrow e^+ e^- \gamma$) as well
as the conversion electron pairs.
%Since we did not apply any vertex position cut in the conversion pair
%reconstruction, this factor, therefore,  included the effect on
%the Dalitz pairs.
In order to estimate the number of remaining background tracks from
$\gamma$-conversions, we multiplied
the number of the reconstructed conversion pairs in the experiment
by $\eta$ in each $P_T$ bin and obtained the cross-hatched histogram in
Fig. \ref{fig:bg}.
The $\gamma$-conversion and Dalitz decay background mainly occupies the
low-$P_T$ region, and comprises 23.8\% of the selected electron sample.

The background from the single-photon annihilation ($e^+e^-
\rightarrow (\gamma ) \rightarrow q\bar{q}$)
was estimated by a Monte-Carlo
simulation with JETSET 6.3\cite{LUND}.
The parameters of this Monte-Carlo program have been
tuned  by fitting
the TOPAZ single-photon annihilation data\cite{adachi}.
We set the maximum
fractional photon energy ($k_{max}$) to be 0.99 for initial-state radiation.
The result is shown in Fig. \ref{fig:bg} as the open area of the histogram.
The electrons which came from single-photon annihilations had high $P_T$ in
general, and were estimated to be 11.8\%.

The contributions from the $e^+e^- \rightarrow e^+e^- \tau^+ \tau^-$
and $e^+e^- \rightarrow \tau^+ \tau^-$ processes
were studied by Monte-Carlo
simulations which were used in previous four-lepton\cite{KUR88,eetutu}
and lepton pair production\cite{tutu} analyses.
The result is shown in Fig.\ref{fig:bg} as the singly-hatched area
of the histogram,
which corresponds to 12.1\% of the  selected electron sample.
This background has a broad $P_T$ distribution.

As for the $e^+e^- \rightarrow e^+e^-e^+e^-(\gamma)$ process,
we checked its contamination experimentally.
First, we searched for events consisting of tracks all of which were
consistent with the electron hypothesis (using dE/dx).
The contamination was only 3.0\%.
The change of the cut on the charged multiplicity in the
event selection from 4 to 5 made only 1.6\% difference seen in the electron
yield. We took these numbers as being systematic errors
in the cross-section evaluation.

The beam-gas background was checked by using off-vertex events.
The electron yield in these events was consistent with zero.

These background contributions were subtracted
from the data on a bin-by-bin basis in
further analysis. In total,
112.3 $\pm$ 15.4(stat.) $\pm$ 11.9(sys.)(anti-tag) and
42.9 $\pm$ 9.3(stat.) $\pm$ 6.2(sys.)(remnant-jet-tag)
electrons remained after the background subtraction.

\section{Results}
\subsection{Inclusive electron cross sections}
In order to compare the data with the theoretical predictions
directly, we carried out acceptance corrections  and measured
(1) the transverse momentum($P_{T}$) dependence of the inclusive
electron cross sections in $\gamma \gamma$ collisions
with the anti-tag condition and (2)
the subprocess cross sections which corresponds to the resolved-photon
processes. As mentioned before, the polar angles of the selected
electrons were
limited to $|\cos\theta| \le 0.77$ in order to ensure good detector
performance in the electron identification.
The acceptance
was determined by the Monte-Carlo simulations of the
direct and resolved processes. The resultant acceptance increased
smoothly from 5\%
to 12\% as the electron-$P_{T}$ goes up from 0.4 GeV to 3.0 GeV.

The results are shown in Figs.\ref{fig:result}-(a) and (b)
for the inclusive electron cross section as a function of
the electron $P_{T}$, and
 in Fig.\ref{fig:fcla} for the cross sections which correspond to the
 resolved-photon processes. Both of the results are also summarized
in Tables \ref{tab:table1}
 and \ref{tab:table2}, where the first errors are statistical and
the second ones are systematic (as discussed below). The errors in
Figs.\ref{fig:result}-(a), (b), and Fig.\ref{fig:fcla}
include both of these errors added in quadrature.

The latter cross sections were obtained by using the subsample of
 remnant-jet-tagged events.
With the Monte-Carlo simulation of the resolved-photon process,
 the efficiency of the remnant-jets leaving an activity greater
than 0.5   GeV in the FCL was evaluated
to be  73$\pm$2\%, where the error  was obtained by
comparing the two cases both
with and without an intrinsic $P_{T}$
in the remnant-jet production.
On the other hand,
that of the direct process was estimated to be about 7\%.
This contribution was subtracted in order to obtain the cross section of
the resolved-photon process.
In spite of an extensive shielding system protecting the FCL
from beam-induced background,
 continuous(random) activities were
still observable. These activities were estimated to be
10\%  from independent event samples(large-angle Bhabha events
and random triggered events) and were subtracted.

The inclusive electron cross section integrated over the
intervals  $ 0.4 \leq P_{T} \leq  3.0$  GeV and
 $|\cos \theta |\leq 0.77$ was measured to be
\begin{center}
 19.3 $\pm$ 2.7 (stat.) $\pm$ 2.1 (sys.) pb,
\end{center}
while that of the subprocess (resolved-photon process) over the same
intervals was
\begin{center}
7.8 $\pm$ 1.7 (stat.) $\pm$ 1.1 (sys.) pb.
\end{center}

\subsection{Systematic errors}
The systematic uncertainty quoted above resulted from the various sources
discussed below.

\begin{enumerate}
\item The uncertainty from the background estimation for
      $\gamma$-conversions and Dalitz decays was estimated
      to be 10 - 18\% by changing the cuts on $V^{0}$ reconstruction,
      and by using independent information from the inner tracking device.
      Since we used the actual number of $V^{0}$'s to estimate the
      background, the statistical error on that number was dominant.
\item The cut dependence was studied by changing the cuts for
      the event  and electron selections by $\pm 10\%$ from the nominal
      cut values.
      The uncertainty estimated this way was about 15\%.
\item Two fitting methods were applied in order to obtain the number
      of electrons from  the dE/dx distribution in each $P_{T}$ bin.
      First, the fitting was carried out by fixing
      the  center value and the resolution to the values obtained from
      the entire electron sample.
      In the second method, the fitting was carried out
      by making these parameters completely free.
      The two methods gave a difference of about 5\%.
\item The uncertainty in the luminosity measurement was 4\%.
\item The Monte-Carlo statistics for the acceptance correction and the
      background estimation for the processes
          $e^{+}\!e^{-}\rightarrow\,e^{+}\!e^{-}
              \!\tau^{+}\!\tau^{-}$
      and $e^{+}\!e^{-}\rightarrow\,q\bar{q}$ were treated
      as systematics errors.
      The total systematic error from the Monte-Carlo statistics was 5\%.
\item As described before, we regarded
      the $e^{+}\!e^{-}\rightarrow\,e^{+}\!e^{-}
      \!e^{+}\!e^{-}$ background as being a systematic error source.
      The systematic error from this source differed from bin to bin,
      but was typically about 4\%.
\item The uncertainty in the tagging efficiency of the remnant-jet (2\%)
      was included in the error on the sub-process cross section.
\end{enumerate}

 The total systematic errors were then obtained to be from 20
 to 35\%, depending on the value of $P_{T}$.

\section{Discussions}

The predictions of the charm-quark-production cross sections
based on the direct and resolved-photon processes are compared with
the data in Figs.\ref{fig:result}-(a) and (b), and the integrated cross
sections are summarized in Table \ref{tab:table2}.
 The lowest-order(LO) predictions are shown in
Fig.\ref{fig:result}-(a), while Fig.\ref{fig:result}-(b) shows the
higher-order(next-to-leading-order(NLO))
 predictions\cite{NLO,enomotoprd,enomotopl}. The NLO corrections increase
the cross sections by about 31\% for the direct process. The effects of
the charm-quark mass are also shown in the two histograms of
direct+LAC1 predictions in Figs. \ref{fig:result}-(a) and (b)
as well as in Table \ref{tab:table2}.
{}From these figures and the table, it is clear that the direct
process(solid histogram) alone falls short of the observed
inclusive electron cross section, being
consistent with the previous $D^{*\pm}$
analysis\cite{enomotoprd,enomotopl}.

The direct process alone cannot explain the remnant-jet activities, either.
The subprocess cross sections obtained from the remnant-jet-tag sample
are compared to the predictions from the resolved-photon processes
alone in Fig. \ref{fig:fcla}, where the NLO corrections are included in the
theoretical predictions.
{}From this figure and Fig. \ref{fig:result}, we can conclude that the
data agree better with the prediction based on the LAC1 parametrization
than that based on the DG parametrization.
As mentioned before, this inclusive electron measurement is sensitive
to the difference in the parton distributions for the kinematical
region of $x$ down to 0.02.
 The difference in the predictions of LAC1 and DG
 results mainly from the difference
  in the gluon distribution at small $x$ less than 0.1.
Our results indicate that the gluon content increases rather rapidly
at the small-$x$ region($x < 0.1$).

\section{Conclusions}
We have studied inclusive electron productions in $\gamma \gamma$
collisions  at $\sqrt{s}=$ 58 GeV and
 measured the inclusive electron cross sections, imposing an
anti-tag condition.  The subprocess cross
sections corresponding to resolved-photon processes were also
measured separately
by using a remnant-jet-tag sample.
The results were compared with the predictions of
charm-quark  production
based on the direct and resolved-photon processes. Here,
in the  predictions, the uncertainties caused by the charm-quark
mass and the effect of the next-to-leading order corrections were studied.
The direct process alone can explain neither the cross sections nor
the remnant-jet activities.
 The comparison  with
 the predictions of the sum of the direct and resolved-photon processes
showed that our data were well explained if the
LAC1 parametrization, which has relatively large gluon contents at
small x,  was taken for the parton distribution in a photon
and the NLO corrections were included.
On the other hand, the DG parametrization could not explain the data even
if the NLO corrections and the effect of the charm-quark mass were
taken into account.

\section*{Ackowledgement}
We appreciate useful discussions with Drs. D. Aurenche, M. Drees,
and K. Hagiwara concerning the cross section of the charm-quark
production and the importance of anti-tag condition.
We thank the TRISTAN accelerator staff
for the successful operation of TRISTAN. We also thank
all of the engineers and technicians at KEK and
the other collaborating institutions: Messers H. Inoue, N. Kimura,
K. Shiino, M. Tanaka, K. Tsukada, N. Ujiie,
and H. Yamaoka.

\newpage
\section*{Table 1, M. Iwasaki et. al., Physics Letters B.}
\begin{table}[h]
\begin{center}
\begin{tabular}{clll}
\hline \hline
electron-$P_{T}$ & &$d\sigma/dP_{T}$ & \\
(GeV) & anti-tag & remnant-jet-tag & no-tag\\ \hline
0.4 - 0.5 & 56.1 $\pm$ 17.1 $\pm$ 13.6 &26.0 $\pm$ 11.4 $\pm$ 7.2 &
            71.3 $\pm$ 17.3 $\pm$ 16.4\\
0.5 - 0.6 & 50.6 $\pm$ 13.8 $\pm$ 10.9 &20.3 $\pm$ 8.3  $\pm$ 5.2&
            48.7 $\pm$ 11.9 $\pm$ 10.6\\
0.6 - 0.8 & 20.8 $\pm$ 6.0 $\pm$ 4.8 &8.0  $\pm$ 3.0  $\pm$ 2.3&
            26.6 $\pm$ 5.9  $\pm$ 5.9\\
0.8 - 1.1 & 8.2 $\pm$ 2.8 $\pm$ 1.7 &2.9  $\pm$ 1.7  $\pm$ 1.1&
            12.7 $\pm$ 2.9  $\pm$ 2.2\\
1.1 - 1.5 & 2.2 $\pm$ 1.3 $\pm$ 1.2 &1.3  $\pm$ 1.1  $\pm$ 0.8&
            3.8  $\pm$ 1.4  $\pm$ 2.0\\
1.5 - 3.0 & 0.7 $\pm$ 0.3 $\pm$ 0.3 &0.2  $\pm$ 0.2  $\pm$ 0.2&
            0.9  $\pm$ 0.3  $\pm$ 0.3\\
\hline \hline
\end{tabular}
\end{center}
\caption{
\baselineskip 20pt
Differential cross sections $d\sigma / dP_{T}$
($| \cos \theta | \leq 0.77$)(pb/GeV) of inclusive electron
productions shown as functions of the electron-$P_{T}$.
The cross sections with anti-tag and no-tag conditions, as well as
the remnant-jet-tag cross section corresponding to the resolved-photon
process are shown in the table,
where the cross section under the no-tag condition was obtained from the
sample without the anti-tag cut (criterion 5) in section 2.1.
This cross section was obtained by using the detection efficiency
calculated with the photon-flux factor of
eq.(1) with $\theta_{max} = \pi$.
}
\label{tab:table1}
\end{table}

\newpage
\section*{Table 2, M. Iwasaki et. al., Physics Letters B.}
\begin{table}[h]
\begin{center}
\begin{tabular}{llcccccccc}
\hline \hline
&&\multicolumn{4}{c}{anti-tag }
&\multicolumn{4}{c}{remnant jet-tag }\\ \hline
$\sigma_{data}$ & (pb) &\multicolumn{4}{c}{ 19.3 $\pm$ 2.7 $\pm$ 2.1 }
 &\multicolumn{4}{c}{ 7.8 $\pm$ 1.7 $\pm$ 1.1 }\\ \hline
& &\multicolumn{2}{c}{LO} &\multicolumn{2}{c}{NLO}
&\multicolumn{2}{c}{LO} &\multicolumn{2}{c}{NLO} \\
charm-quark mass & (GeV)  & 1.5 & 1.3 & 1.5 & 1.3 &
 1.5 & 1.3 & 1.5 & 1.3 \\ \hline
$\sigma_{direct}$ &     & 5.8 & 6.7 & 7.6 & 8.8 &  - & -& - & -\\
$\sigma_{DG}$     &(pb) & 1.7 & 2.2 & 2.0 & 2.5 & 1.7 & 2.2 &2.0 & 2.5\\
$\sigma_{LAC1}$    &    & 5.0 & 6.8 & 6.3 & 8.3 & 5.0 & 6.8 &6.3 & 8.3\\ \hline
$\sigma_{direct+DG}$   && 7.5 & 8.9 & 9.6 & 11.3& - & - & - & -\\
$\sigma_{direct+LAC1}$ &(pb)& 10.8& 13.5& 13.9& 17.1& - & - & - & -\\ \hline
\hline
\end{tabular}
\end{center}
\caption{
\baselineskip 20pt
Measured inclusive electron cross sections integrated over the
$P_{T}$ region from 0.4 to 3.0 GeV for $|\cos \theta
|\leq 0.77$, together with the theoretical predictions
based on the direct and resolved-photon processes.
The dependence on the charm-quark mass and the effect of the NLO
corrections are also shown.}
\label{tab:table2}
\end{table}
\newpage
\section*{Figure captions}
\begin{figure}[h]
\caption{
\baselineskip 20pt
Examples of diagrams contributing to the hadron production in two-photon
processes: (a) direct(QPM) process,(b) one-resolved-photon process,
(c) two-resolved process, (d) VDM process.
}
\label{fig1}
\caption{
\baselineskip 20pt
Various distributions for electron candidates:
(a) the distance between a TPC track and
its closest BCL cluster,
(b) E/P, (c) the confidence level (CL) for the TOF, and (d) dE/dx.
Each of these distributions was obtained
with all the cuts but that on the plotted quantity.
}
\label{fig:selection}
\caption{
\baselineskip 20pt
Observed number of electrons as a function of $P_T$.
The closed circles are
the experimental data, while the histograms are the background components
estimated by various methods described in the text.
The cross-hatched area is
the background from $\gamma$-conversions and $\pi^0$ Dalitz decays,
the singly-hatched area is from
$e^{+}\!e^{-}\rightarrow\,e^{+}\!e^{-}\!\tau^{+}\!\tau^{-}$, and the open area
from single photon annihilations.
}
\label{fig:bg}
\caption{
\baselineskip 20pt
Differential cross section, $d\sigma / dP_{T}$
($|\cos \theta |\leq 0.77$)(pb/GeV), of inclusive electron productions plotted
against the electron-$P_T$, where anti-tag condition is imposed.
The closed circles
are the experimental data, while the histograms are the theoretical
predictions of
(a) LO and (b) NLO ; the solid lines are the direct only,
the dotdashed lines are the direct+DG, the dotted lines are the direct+LAC1
with a charm-quark mass of 1.5GeV,
and the dashed lines are the direct+LAC1 with a charm-quark mass of 1.3GeV.
}
\label{fig:result}
\end{figure}
\begin{figure}[h]
\caption{
\baselineskip 20pt
Differential cross section, $d\sigma / dP_{T}$
($|\cos \theta |\leq 0.77$)(pb/GeV), of inclusive electron productions
as a function of the electron-$P_T$ obtained from
the remnant-jet-tagged sample.
The closed circles are the experimental data, while
the histograms are the theoretical predictions with the NLO corrections;
the dotdashed line is DG,
the dotted line is LAC1 with a charm-quark mass of 1.5GeV, and
the dashed line is LAC1 with a charm-quark mass of 1.3GeV.
}
\label{fig:fcla}
\end{figure}
\newpage
\section*{Figure 1}
\epsfysize16cm
\hskip-0.5in
\epsfbox{fig1.epsf}
\newpage
\section*{Figure 2}
\epsfysize18cm
\hskip-0.5in
\epsfbox{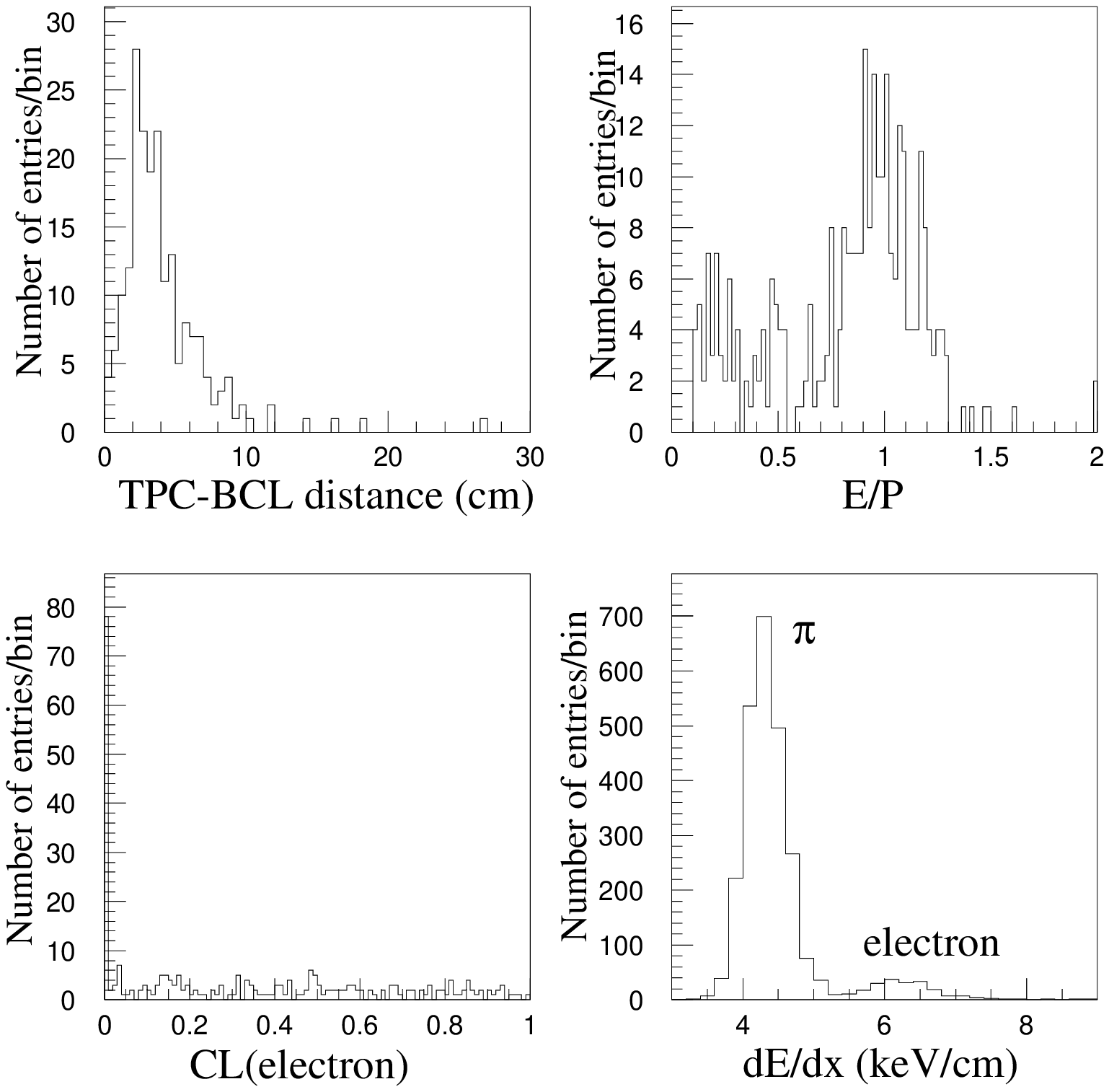}
\newpage
\section*{Figure 3}
\epsfysize9cm
\epsfbox{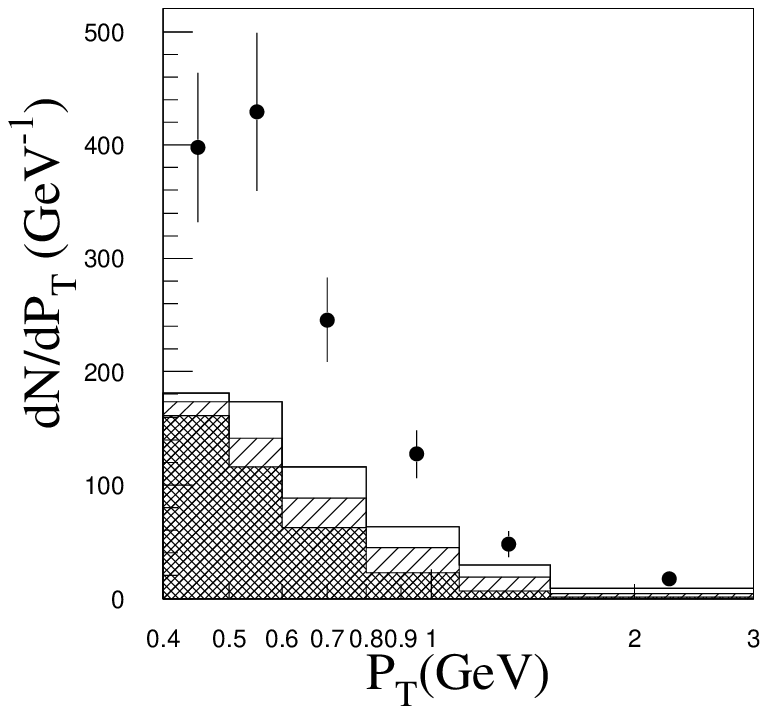}
\hskip-0.5in
\newpage
\section*{Figure 4}
\epsfysize9cm
\hskip-0.5in
\epsfbox{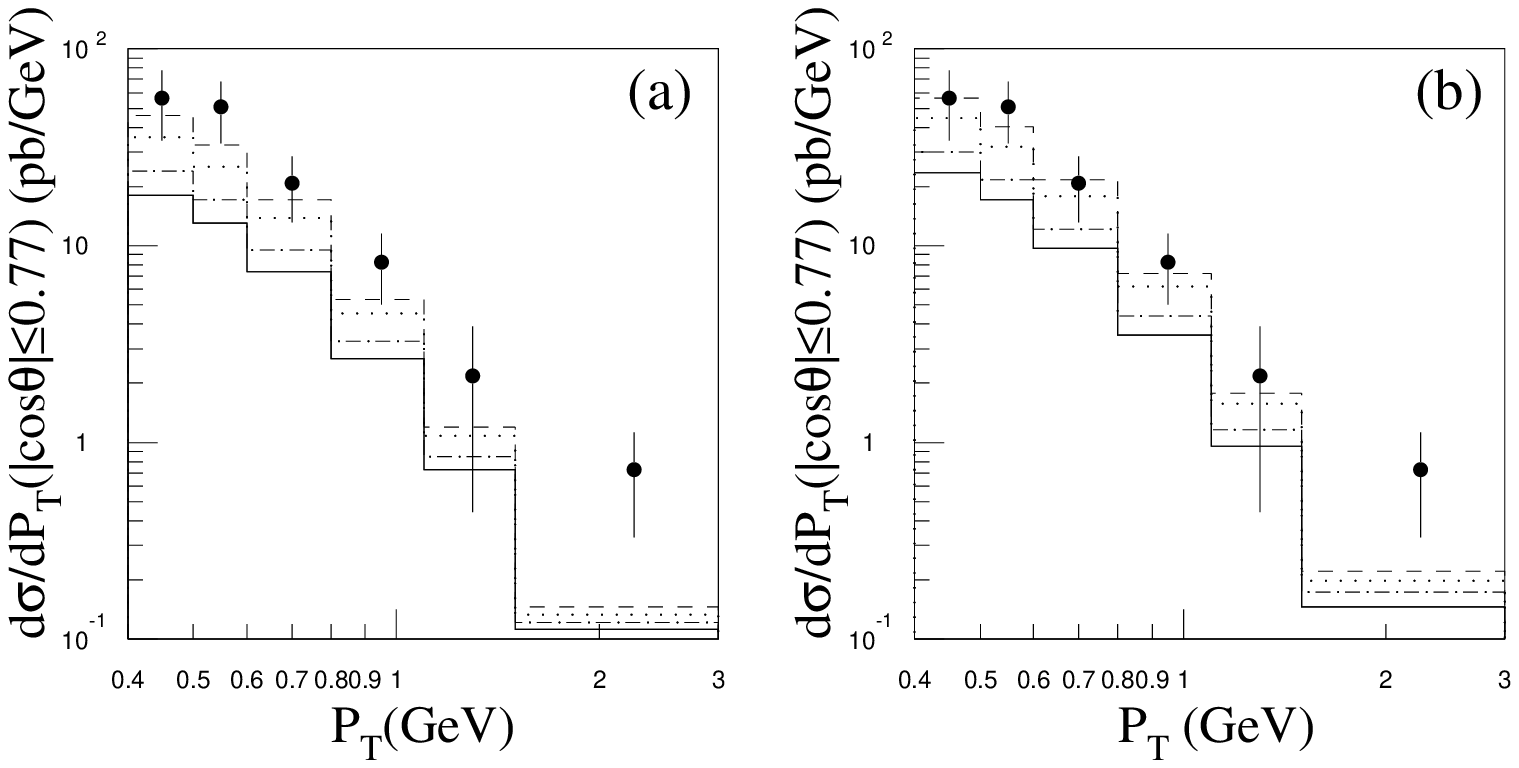}
\newpage
\section*{Figure 5}
\epsfysize9cm
\hskip-0.5in
\epsfbox{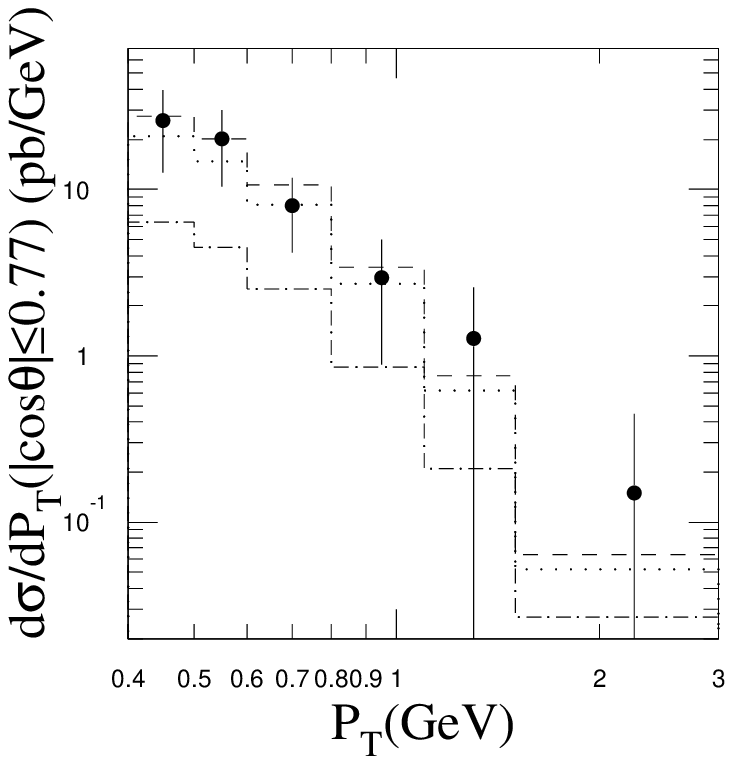}

\newpage
\begin{thebibliography}{99}
\bibitem{PEPPETRA}
PLUTO Collab., Ch. Berger {\it et. al.},
Z. Phys. {\bf C29} (1985) 499; Z. Phys. {\bf C33} (1987) 351; \\
TPC/2$\gamma$ Collab.,
H. Aihara {\it et.al.},Phys. ReV. {\bf D41} (1990) 2667; \\
CELLO Collab.,
H. j. Behrend {\it et.al.}, Z. Phys. {\bf C51} (1991) 365.
\bibitem{HAYASHII}
TOPAZ Collab., H. Hayashii {\it et.al.},
Phys. Lett. {\bf B314} (1993) 149.
\bibitem{AMY}
AMY Collab., R. Tanaka {it et.al.},
 Phys. Lett. {\bf B277} (1992) 215;
B.J. Kim {\it et.al.}, Phys.Lett. {\bf B325} (1994) 248.
\bibitem{ALEPH}
ALEPH Collab., D. Buskulic {\it et.al.},
Phys. Lett. {\bf B313} (1993) 509.
\bibitem{DELPHI}
DELPHI Collab., P. Abreu  {\it et.al.}, CERN-preprint, CERN-PPE/94-04.
\bibitem{DG90}
M. Drees and R. M. Godbole, Nucl. Phys. {\bf B339} (1990) 355.
\bibitem{AU93}
P. Aurenche {\it et.al.}, KEK-preprint, KEK-93-180.
\bibitem{BR78}
S. J. Broadsky, T. A. DeGrand, J. F. Gunion, and J. H. Weis,
Phys. Rev. Lett. {\bf 41} (1978) 672,
and  Phys. Rev. {\bf D19} (1979) 1418;
G. H. Llewellyn-Smith, Phys. Lett. {\bf 79B} (1978) 83;
H. Terazawa, J. Phys. Soc. of Japan, {\bf 47} (1979) 355;
K. Kajantie and R. Raitio, Nucl. Phys. {\bf B159} (1979) 528.
\bibitem{H1ZEUS}
H1 Collab., T. Ahmed {\it et.al.}, Phys. Lett. {\bf B297} (1992) 205; \\
ZEUS Collab., M. Derrick {\it et.al.}, Phys. Lett. {\bf B297} (1992) 404.
\bibitem{LAC}
H. Abramowicz, K. Charchula, and A. Levy,
Phys. Lett. {\bf B269} (1991) 458.
\bibitem{NLO}
M. Drees, M. K\"{a}mer, J. Zunft, and P. M. Zerwas,
Phys. Lett. {\bf B306} (1993) 371;
J. Smith and W. van. Neerven, Nucl. Phys. {\bf 274} (1992) 36.
\bibitem{VDM}
J. J. Sakurai and D. Schildknecht, Phys. Lett. {\bf B40} (1979) 121;
I. F. Ginzburg and V. G. Selbo, Phys. Lett. {\bf B109} (1982) 231.
\bibitem{enomotoprd}
TOPAZ Collab., R. Enomoto et. al., Phys. Rev. {\bf D50} (1994) 1879.
\bibitem{enomotopl}
TOPAZ Collab., R. Enomoto et. al., Phys. Lett. {\bf B328} (1994) 535.
\bibitem{CPEP}
TASSO Collab., W. Braunschweig {\it et.al.},
Z. Phys. {\bf C47} (1990) 499; \\
TPC/$2\gamma$ Collab. M. Alston-Garnjost {\it et.al.},
Phys. Lett. {\bf B252} (1990) 499.
%\bibitem{DO}
%D. W. Duke and J. F. Owens,  Phys. Rev. {\bf D22} (1980) 2280.
\bibitem{DG}
M. Drees and K. Grassie, Z. Phys. {\bf C28} (1985) 451.
\bibitem{GRV}
M. Gl\"{u}ck, E. Reya, and A. Vogt, Phys. Rev. {\bf D46} (1992) 1973.
\bibitem{AU92}
P. Aurenche, P. Chiappeta, M. Fontannaz, J. P. Guillet and E. Pilon,
Z. Phys. {\bf C56} (1992) 589.
\bibitem{GS}
L. E. Gordon and J. K. Storrow, Z. Phys., {\bf C56} (1992) 307.
\bibitem{HAG93}
H. Hagiwara, M. Tanaka, I. Watanabe, and T. Izubuchi,
KEK-preprint, KEK-93-160.
\bibitem{VENUS}
VENUS Collab., S. Uehara et. al., Z. Phys. {\bf C63} (1994) 213.
\bibitem{EPA}
C. Weiz\"acker, Z. Phys. {\bf 88} (1934) 612;
E. J. Williams, Phys. Rev. {\bf 45} (1934) 729.
\bibitem{HAG91}
K. Hagiwara, H. Iwasaki, A. Miyamoto, H. Murayama and
D. Zeppenfeld, Nucl. Phys. {\bf B365} (1991) 544.
\bibitem{flux}
S. Frixione, M. L. Mangano, P. Nason, and G. Ridolfi
Phys. Lett. {\bf B319} (1993) 339.
\bibitem{KUR88}
M. Kuroda, Meiji Gakuin University (Tokyo) Research J. 424 (1988)27.
\bibitem{BASES}
S. Kawabata, Comp. Phys. Comm., {\bf 41} (1986) 127.
\bibitem{LUND}
T. Sj\"{o}strand, Comp. phys. Comm., {\bf 39} (1986) 347;
T.Sj\"{o}strand and M.Bengtsson, Comp. phys. Comm., {\bf 43} (1987) 367.
%\bibitem{pythia} T. Sj\"ostrand, CERN-TH.6488/92, unpublished.
\bibitem{TOPAZ}
TPC: T. Kamae {\it et. al.}, Nucl. Instrum. Meth. {\bf A252} (1986) 423;
Masking: T. Kishida {\it et. al.}, Nucl. Instrum. Meth. {\bf A254} (1987) 367;
BCL: S. Kawabata {\it et. al.}, Nucl. Instrum. Meth. {\bf A270} (1988) 11;
ECL: K. Fujii {\it et. al.}, Nucl. Instrum. Meth. {\bf A236} (1985) 55;
J. Fujimoto, {\it et. al.}, Nucl. Instrum. Meth. {\bf A256} (1987) 449.
\bibitem{FCL}
FCL: H.Hayashii {\it et. al.}, Nucl. Instrum. Meth. {\bf A316} (1992) 202.
\bibitem{KI93}
Masking: H. Kichimi {\it et.al.}, Nucl. Instrum. Meth. {\bf A334} (1993) 367.
\bibitem{trig}
Trigger: R. Enomoto {\it et. al.},
Nucl. Instrum. Meth. {\bf A269} (1988) 507;
R. Enomoto, K. Tsukada, N. Ujiie and A. Shirahashi,
IEEE Trans. NS. {\bf Vol. 35, No. 1}, 419 (1988);
T. Tsukamoto, M. Yamauchi and R. Enomoto,
Nucl. Instrum. Meth. {\bf A297} (1990) 148.
\bibitem{TAU94}
T. Tauchi, KEK-preprint, KEK-94-67, to be appeared in the proceedings of
XXIXth Recontres de Moriond ``QCD and High Energy Interactions'',
Meribel, France, March 19-26, 1994.
\bibitem{nagai}
TOPAZ collab., K. Nagai {\it et. al.}, Phys. Lett. {\bf B278} (1992) 506.
\bibitem{adachi}
TOPAZ collab., I. Adachi {\it et. al.}, Phys. Lett. {\bf B227} (1989) 495.
\bibitem{eetutu}
TOPAZ collab., H. Hayashii {\it et. al.}, Phys. Lett. {\bf B279} (1992) 422.
\bibitem{tutu}
TOPAZ collab., B. Howell {\it et. al.}, Phys. Lett. {\bf B291} (1992) 206.
\end{thebibliography}
\end{document}